\documentclass[trackchanges,twocolumn,tighten]{aastex62} 

\usepackage{amsmath}

\usepackage{longtable}
\usepackage{soul}
\usepackage{CJK}
\usepackage{lineno}

\newcommand{\bbud}{\alpha\;\mathrm{Ori~B}}

\definecolor{meridithgreen}{RGB}{0, 150, 0}
\definecolor{jaredpurple}{RGB}{93, 63, 211}
\definecolor{annablue}{RGB}{29, 162, 219}

\newcommand{\appropto}{\mathrel{\vcenter{
		\offinterlineskip\halign{\hfil$##$\cr
	\propto\cr\noalign{\kern2pt}\sim\cr\noalign{\kern-2pt}}}}}

\newlength{\apjcolwidth}
\newlength{\figwidth}
\newlength{\doublewide}
\begin{document}
\title{Betelgeuse’s Buddy: X-Ray Constraints on the Nature of $\alpha$ Ori B}

\author[0000-0002-7296-6547]{Anna J.G. O'Grady}
\altaffiliation{Contributed Equally}
\altaffiliation{McWilliams Fellow}
\affiliation{McWilliams Center for Cosmology and Astrophysics, Department of Physics, Carnegie Mellon University, Pittsburgh, PA 15213, USA}

\author[0000-0002-9700-0036]{Brendan O'Connor}
\altaffiliation{Contributed Equally} 
\altaffiliation{McWilliams Fellow}
\affiliation{McWilliams Center for Cosmology and Astrophysics, Department of Physics, Carnegie Mellon University, Pittsburgh, PA 15213, USA}

 \author[0000-0003-1012-3031]{Jared A. Goldberg}
 \affiliation{Center for Computational Astrophysics, Flatiron Institute, New York, NY, USA}

 \author[0000-0002-8717-127X]{Meridith Joyce}
 \affiliation{University of Wyoming, 1000 E University Ave, Laramie, WY USA}
 \affiliation{Konkoly Observatory, HUN-REN CSFK, Konkoly-Thege Mikl\'os \'ut 15-17, H-1121, Budapest, Hungary}
 \affiliation{CSFK, MTA Centre of Excellence, Konkoly-Thege Mikl\'os \'ut 15-17, H-1121, Budapest, Hungary}

 \author[0000-0002-8159-1599]{L\'{a}szl\'{o} Moln\'{a}r}
 \affiliation{Konkoly Observatory, HUN-REN CSFK, Konkoly-Thege Mikl\'os \'ut 15-17, H-1121, Budapest, Hungary}
 \affiliation{CSFK, MTA Centre of Excellence, Konkoly-Thege Mikl\'os \'ut 15-17, H-1121, Budapest, Hungary}
 \affiliation{E\"otv\"os Lor\'and University, Institute of Physics and Astronomy, P\'azm\'any P\'eter s\'et\'any 1/A, H-1117, Budapest, Hungary}

\author[0000-0002-8878-3315]{Christian I. Johnson}
\affiliation{Space Telescope Science Institute
3700 San Martin Drive
Baltimore, MD 21218, USA}

 \author[0000-0002-8548-482X]{Jeremy Hare}
    \affiliation{Astrophysics Science Division, NASA Goddard Space Flight Center, 8800 Greenbelt Rd, Greenbelt, MD 20771, USA}
    \affiliation{Center for Research and Exploration in Space Science and Technology, NASA/GSFC, Greenbelt, Maryland 20771, USA}
    \affiliation{The Catholic University of America, 620 Michigan Ave., N.E. Washington, DC 20064, USA}

\author[0000-0001-5228-6598]{Katelyn Breivik}
\affiliation{McWilliams Center for Cosmology and Astrophysics, Department of Physics, Carnegie Mellon University, Pittsburgh, PA 15213, USA}

\author[0000-0001-7081-0082]{Maria R. Drout}
\affiliation{David A. Dunlap Department of Astronomy \& Astrophysics, University of Toronto, 50 St. George Street, Toronto, Ontario, M5S 3H4, Canada}

\author[0000-0002-0870-6388]{Maxwell Moe}
 \affiliation{University of Wyoming, 1000 E University Ave, Laramie, WY USA}

 \author[0000-0002-0882-7702]{Annalisa Calamida}
 \affiliation{Space Telescope Science Institute, 3600 San Martin Drive, Baltimore, 21218 MD, USA}
 \affiliation{INAF - Osservatorio Astronomico Capodimonte, Salita Moiariello 16, 80131 Napoli, Italy}

\correspondingauthor{Anna O'Grady}
\email{aogrady@andrew.cmu.edu}

\begin{abstract}

The $\sim$\,$2100$\,d Long Secondary Period of Betelgeuse's optical lightcurve and radial velocity motivated the prediction of a low-mass stellar companion, expected to be at maximal apparent separation from Betelgeuse around December 2024. We carried out Director's Discretionary Time observations with the \textit{Chandra X-ray Observatory} to identify any X-ray emission from the companion and constrain its nature as either a compact object or young stellar object (YSO). Past X-ray observations occurred at the wrong phase of the companion's orbit for optimal detection prospects and/or lacked the deep exposure required to constrain the typical X-ray luminosities of YSOs. In our 41.85 ks exposure with \textit{Chandra}, we do not detect an X-ray source at the position of Betelgeuse. For an estimated hydrogen column density $N_H$\,$=$\,$6\times10^{22}$ cm$^{-2}$, we place a limit on the X-ray luminosity of $L_X$\,$\lesssim$\,$2\times10^{30}$ erg s$^{-1}$ ($\lesssim$\,$4.7\times10^{-4}L_\odot$) in $0.5$\,$-$\,$8$ keV for a 10 MK plasma temperature spectral model, or $L_X$\,$\lesssim$\,$5\times10^{29}$ erg s$^{-1}$ ($\lesssim$\,$1.2\times10^{-4}L_\odot$) for an absorbed power law with photon index $\Gamma$\,$=$\,$2$. These limits robustly exclude an accreting compact object (white dwarf or neutron star) as the companion. Solar mass YSOs with an age similar to Betelgeuse ($\sim$10 Myr) display a range of X-ray luminosities ($10^{28-32}$ erg s$^{-1}$), and we can place upper bounds within this range for most absorbing columns. Based on these considerations, we conclude that the companion to Betelgeuse is likely a low-mass YSO.

\end{abstract}


\section{Introduction}
\label{sec:intro}

Betelgeuse (or $\alpha$ Orionis), the closest red supergiant (RSG) to the Sun ($\sim$168 pc, \citealt{Joyce2020}), has been extensively studied through the history of astronomy\footnote{See, e.g., an analysis of historical records which suggests that the star had a significant shift towards redder colors over a few millennia, experiencing a rapid phase of stellar evolution within human memory \citep{Neuhauser2022}.}. Betelgeuse was the first star beyond the Sun whose surface was directly imaged, revealing large convective hot spots  \citep{Gilliland-1996,Haubois-2009,o'gorman-2017}. Molecular layers and dust shells surrounding the star were mapped with interferometric measurements \citep[e.g.][]{Ohnaka-2009,Haubois2019}. The star also shows photometric variability on tens-to-thousands of day timescales from convection, radial pulsations, and dust formation \citep[e.g.][]{Kiss2006,Joyce2020,Jadlovsky2023,MacLeod2023,MacLeod2024}. The recent Great Dimming of late 2019/early 2020 sparked many studies aimed at explaining the sudden decrease in brightness, motivated by data collected by a wide range of instruments \citep[see, e.g.,][]{Dupree2020,Dupree2022,Kashyap-2020,Levesque2020,Harper-SOFIA-2020,Harper2020,Montarges2021,Wheeler2023}. Although some hoped that this event might mark an imminent supernova explosion \citep[e.g.][]{Saio2023}, resolved imaging revealed that the Great Dimming was caused by a dense dust cloud partially obscuring the disk of the star from our vantage point \citep{Montarges2021}. 
This discovery once again highlighted the importance of the formation and structure of circumstellar material surrounding red supergiant stars \citep{Dupree2022,Cannon-2023, MacLeod2023}.

Recently, two independent analyses posited a companion to Betelgeuse -- $\alpha$~Ori~B, or ``the companion'' in this work. 
\citet{Goldberg2024} demonstrate that the long secondary period (LSP) of Betelgeuse, a $\sim$6 year periodicity in the light curve, coupled with radial velocity (RV) measurements of Betelgeuse, can best be explained by a $\approx$ 1.17$\pm$0.07 M$_{\odot}$ (1-$\sigma$ uncertainty) companion at an orbital separation of $\approx$ 1800 $R_\odot$ ($\approx$ 2.5 Betelgeuse radii). The companion passes in front of Betelgeuse near the maximum-luminosity phase of the LSP, and interacts with the circumstellar environment (see \citealt{Goldberg2024} Figure 4). Likewise, \citet{MacLeod2024} argue for the existence of a companion to Betelgeuse on the basis of astrometric and longer-baseline RV measurements. With mass estimates of $\approx$ 0.6$\pm$0.14 M$_{\odot}$ from the RV (due to a smaller inferred RV amplitude when fitting to the longer dataset including archival data) and $\sim$\,2.2$\pm$0.5 M$_{\odot}$ (1-$\sigma$ uncertainties) inferred from the astrometric variation. Both studies prefer a low eccentricity or near-circular orbit, though neither work puts direct constraints on the eccentricity. These mass estimates are broadly in agreement with each other, as the inference hinges on the amplitude of fluctuations in the star's RV and/or astrometric position, which is sensitive to the sparseness and quality of old data, as well as the temporal baseline of the observations at a factor-of-2 level. Nonetheless, the different estimates for the companion mass would imply different evolutionary pathways for the system, and necessitate independent constraints on the nature of the companion.

Several identities for the companion fit within the plausible mass range of $\sim$\,0.5--2 M$_{\odot}$, including a compact objects -- specifically either a neutron star (NS) or a white dwarf (WD) -- or a standard stellar object. We note that smaller objects, such as brown dwarfs or large Jupiter-class planets, are ruled out by the \citet{MacLeod2024,Goldberg2024} mass constraints. In the stellar object case, given Betelgeuse's approximate age of $\sim$\,10 Myr \citep{Joyce2020} and assuming a co-eval formation channel, the companion would be a pre-main sequence young stellar object (YSO), except in the marginal case of a 2M$_{\odot}$ companion  -- though we note that uncertainties in the lengths of YSO tracks are large \citep{Tognelli2011yso,Baraffe2015yso,Haemmerle2019yso,Amard2019yso}. 

For either a compact object or YSO companion, X-ray emission may be expected. Given the orbital separation, the companion is embedded within the dusty shell surrounding Betelgeuse \citep[e.g.,][]{Haubois2019,Haubois2023}, and would interact with Betelgeuse's wind.  In the compact object scenario (for either a NS or a WD), accretion would therefore be expected \citep{Aizu1973,Livio1986,Blodin1990,Mukai2017,Maldonado2025}, and would lead to a bright X-ray signature. Typical X-ray luminosities for accreting white dwarfs are $10^{30-34}$ erg s$^{-1}$ \citep{Luna2013}, and for accreting neutron stars are $10^{32-36}$ erg s$^{-1}$ \citep{Yungelson2019,De2022a,De2022b}.

In addition, young YSOs are X-ray bright due to strong magnetic dynamo processes leading to coronal ejections and flares \citep{Feigelson1999Review,Preibisch2005,Preibisch2005Review,Getman2022}. The X-ray luminosity is strongly dependent on age and rotation -- younger YSOs rotate faster and are highly active. At the approximate age of Betelgeuse, typical X-ray luminosities for YSOs are $\sim10^{29}-10^{31.5}$ erg/s \citep{Getman2005COUP0,Getman2022}.

From Betelgeuse's RV period and phase information, \citet{Goldberg2024} predicted that the companion and Betelgeuse would be at a maximum apparent separation (also known as quadrature) near 6th December 2024, which is within $\approx$ a month of estimates from the period and phase recovered by \citet{MacLeod2024}. After that date, the putative companion will enter the receding phase of its $\approx6$yr orbit, moving from the observer and behind $\alpha$~Ori, to return to approaching quadrature in late 2027.
Given this, we obtained \emph{Chandra X-ray Observatory} observations of the Betelgeuse system in order to put constraints on the X-ray luminosity of the companion. We also obtained complementary \emph{Hubble Space Telescope} UV spectroscopic observations, presented in a companion paper \citep{BetelHST2025}. In this work, we present our new \textit{Chandra} observations in \S\ref{sec:observations}, which represent the deepest X-ray observations of Betelgeuse to date. We then discuss in \S\ref{sec:results} the implications of these observation vis a vis the stellar properties of the companion. We conclude in \S\ref{sec:conclusions}.

\section{\textit{Chandra} Observations and Analysis}\label{sec:observations}

\begin{figure*}
   \centering
  \includegraphics[width=2\columnwidth]{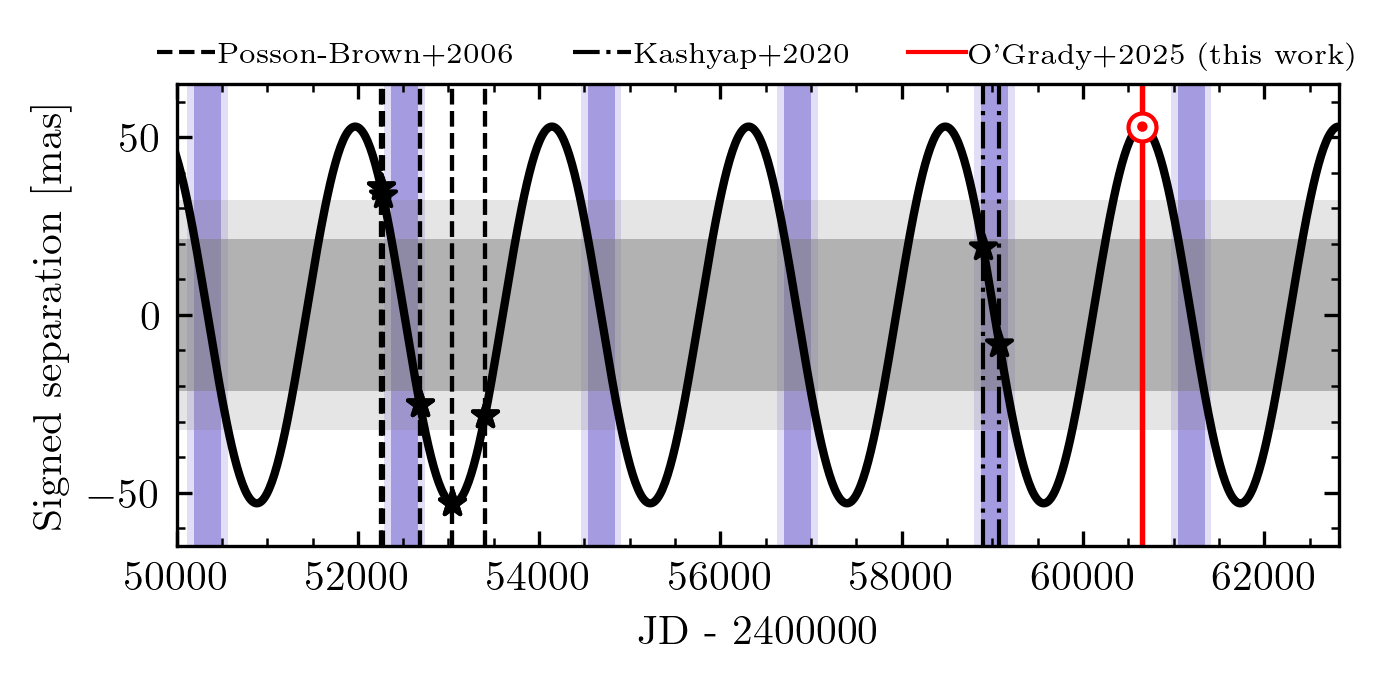}
      \caption{ 
      Separation of the putative companion from Betelgeuse as a function of time using the \citet{Goldberg2024} ephemeris, where the sign represents the  companion moving away from (+) or towards (-) the observer. Vertical lines indicate the relative timing of past on-axis \textit{Chandra} observations (\citealt{Posson-Brown-2006,Kashyap-2020},  black dashed and dot-dashed lines), and our latest observations (red line), which occurred at a previously unprobed phase of the LSP with maximal separation from Betelgeuse. The gray shaded bar represents the 43mas radial extent of Betelgeuse (dark gray) and the $\approx$1.5$R_*$ dust shell inferred by \citep[light gray;][]{Haubois2019,Haubois2023}. Purple shaded regions indicate epochs when the companion is fully obscured by Betelgeuse and the circumstellar shell (dark/light purple). Note that the recent \citet{Kashyap-2020} observations, which took place during the Great Dimming of 2020, occurred while $\alpha$~Ori~B was fully obscured by Betelgeuse, and the \citet{Posson-Brown-2006} observations were strongly off-axis (see text).}
    \label{fig:timing}
\end{figure*}

\subsection{Archival Observations}
\label{sec: archival}

Due to the extreme visual brightness of Betelgeuse ($V$\,$\approx$\,$0.46$ mag), many X-ray telescopes, such as \textit{Swift} \citep[e.g.,][]{Hare2021-swiftOL} or \textit{eROSITA}\footnote{\url{https://erosita.mpe.mpg.de/dr1/erodat/catalogue/source_view/DR1_Main/em01_089084_020_ML00001_002_c010}}, suffer from extreme optical loading\footnote{\url{https://www.swift.ac.uk/analysis/xrt/optical_loading.php}}, making observations of Betelgeuse, or the companion, impossible. However, the instruments mounted on the \textit{Chandra X-ray Observatory} have high quality optical filters that largely prevent optical loading \citep[see, e.g.,][for a discussion]{Posson-Brown-2006}.
The location of Betelgeuse was observed regularly by  \textit{Chandra} between 2001-12-27 and 2007-08-16 for a total of 58 ks as part of calibration observations\footnote{\url{https://cxc.harvard.edu/ccw/proceedings/05_proc/presentations/possonbrown/}}. These observations are shallow $1$\,$-$\,$2$ ks visits and regularly placed Betelgeuse at large off-axis angles from the detector aimpoint, where they are significantly less sensitive to any X-ray photons coming from Betelgeuse or its companion. 
\citet{Posson-Brown-2006} present an analysis of all $\sim$21 ks of on-axis \textit{Chandra} observations occurring through this calibration campaign, including data obtained with ACIS-I, HRC-I, and HRC-S. 
Additional on-axis observations were obtained by \citet{Kashyap-2020} with HRC-I during the period of the Great Dimming on 2020-02-17 (5 ks) and 2020-08-15 (5ks). All of these previous \textit{Chandra} observations resulted in non-detections. 

In Figure \ref{fig:timing} we show the location of the companion of Betelgeuse with respect to the timing of these previous \textit{Chandra} observations. In particular, we find that all past on-axis \textit{Chandra} observations \citep{Kashyap-2020} occurred at the wrong phase of the companion's orbit (e.g., with the companion either totally or partially eclipsed by Betelgeuse), and, in addition, were too shallow. Therefore, we obtained new \textit{Chandra} observations at the optimal orbital phase. Our new observations and analysis are presented below in \S \ref{sec: chandraanalysis}.

\subsection{New Observations with \textit{Chandra}}
\label{sec: chandraanalysis}

\begin{figure*}
   \centering
  \includegraphics[width=2\columnwidth]{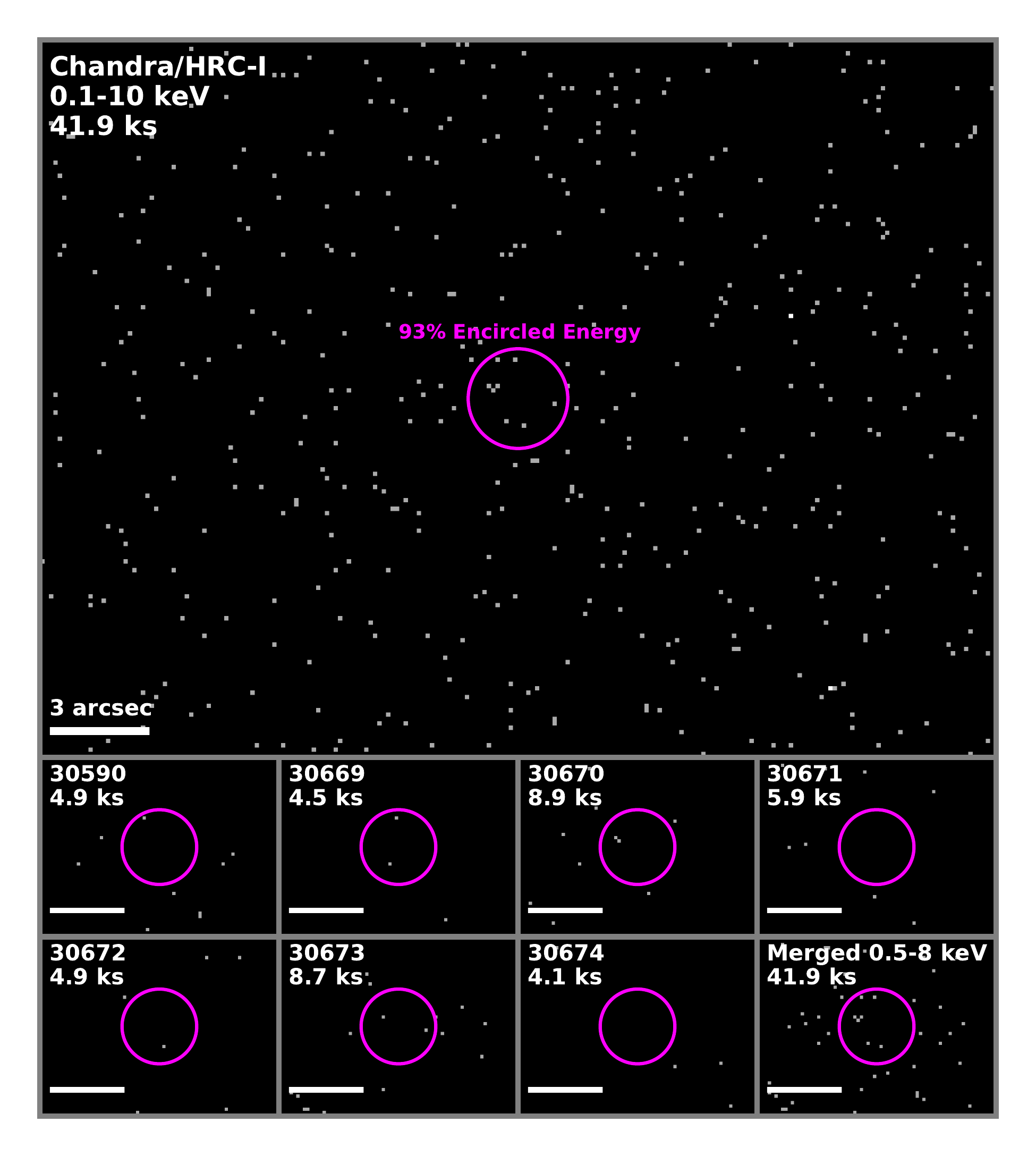}
      \caption{Finding chart showing our observations of Betelgeuse. The top panel displays the merged 41.85 ks image in the full $0.1$\,$-$\,$10$ keV energy range. The bottom panels display individual exposures, and the bottom right shows the merged 41.85 ks image filtered to $0.5$\,$-$\,$8$ keV. The solid magenta circle in all images corresponds to a circular source region with radius 1.5\arcsec, corresponding to the 93\% encircled energy fraction of the PSF. The image is shown in the native 0.13\arcsec\, pixel scale of HRC-I. In all images the orientation is such that North is up and East is to the left. 
              }
    \label{fig:fc}
\end{figure*}

We carried out \textit{Chandra} Director's Discretionary Time (DDT) observations (PI: A. O'Grady; Program: 25209014\footnote{\url{https://doi.org/10.25574/cdc.424}}) of Betelgeuse between 2024-12-09 and 2024-12-15 using the High Resolution Camera (HRC-I) for a total of 41.85 ks. The observations were broken into seven different epochs, see Table \ref{tab: observationsXray} for details. The \textit{Chandra} data were retrieved from the \textit{Chandra} Data Archive (CDA)\footnote{\url{https://cda.harvard.edu/chaser/}}.We re-processed the data using the \texttt{CIAO v4.17.0} data reduction package with \texttt{CALDB v4.11.6} \citep[see][]{Ciao}. We first ran the \texttt{chandra\_repro} command on each individual epoch to remove bad pixels and de-streak the data. We searched each epoch for periods of background solar flares, but found none. We then performed source detection with \texttt{wavdetect}, finding three sources clearly detected in each epoch, albeit at large off-axis angles of $\sim$6\arcmin. Due to their large off-axis angles, leading to an extended point spread function (PSF) and poorly localized centroid ($\sim$1-2\arcsec\, in both RA and DEC at 68\% confidence), we cannot accurately correct the native astrometry of our observations. The localization of each source as computed with \texttt{wavdetect} is consistent with slightly higher quality localizations in the \textit{Chandra} Source Catalog (CSC) version 2.1  \citep[CSC2.1;][]{Evans2024csc}. The offset is consistent with the expectations for the absolute astrometric error of HRC-I\footnote{\url{https://cxc.cfa.harvard.edu/cal/ASPECT/celmon/}} which is accurate to 0.85\arcsec\, at the 90\% confidence level (CL). 

To combine the seven epochs, we ran the \texttt{merge\_obs} command on the full $0.1$\,$-$\,$10$ keV HRC-I energy range. We display the combined image, as well as the individual epochs, in Figure \ref{fig:fc}. There is no obvious source detected at the proper--motion--corrected location of Betelgeuse \citep[e.g.,][]{Harper2017}. In order to quantify this, we define a circular source region of radius 1.5\arcsec\, (93\% encircled energy fraction at 1.5 keV), and an annular background region with inner radius 30\arcsec\, and outer radius of 60\arcsec. The encircled energy fraction was determined using the \texttt{mkpsfmap} tool, which also finds a 90\% encircled energy radius of 0.95\arcsec, computed at 1.5 keV, which is the effective energy of the \textit{Chandra}/HRC-I ``wide'' band. We adopt the larger 1.5\arcsec\, radius given the likely astrometric error is the same order as the 90\% radius. In the 1.5\arcsec\, source aperture, we find a total of 8 total counts in the source region with 4.9 expected background counts. Using the \texttt{aplimits} tool \citep{Kashyap2010}, we estimate a $3\sigma$ count rate upper limit of $<3.7\times10^{-4}$ cts s$^{-1}$ ($0.1$\,$-$\,$10$ keV), where we have applied a standard aperture correction.
We note that our choice of source aperture does not have a strong impact on our results, and using a 0.95\arcsec\,radius source region leads to limits that are deeper by only 20\% (e.g., $<3.0\times10^{-4}$ cts s$^{-1}$ in $0.1$\,$-$\,$10$ keV). 

We further filtered the energy range of the individual event files for each  epoch in order to remove background photons\footnote{\url{https://cxc.cfa.harvard.edu/ciao/threads/hrci_bg_spectra/}}. We selected the PI range $50$\,$-$\,$800$, corresponding to approximately $0.5$\,$-$\,$8$ keV. Summing the source and background counts between the seven epochs, we find 7 total counts in the source region with 3.9 background counts. The corresponding $3\sigma$ upper limit on the count rate is $<3.6\times10^{-4}$ cts s$^{-1}$ ($0.5$\,$-$\,$8$ keV). 

\begin{figure}
   \centering
    \includegraphics[width=\columnwidth]{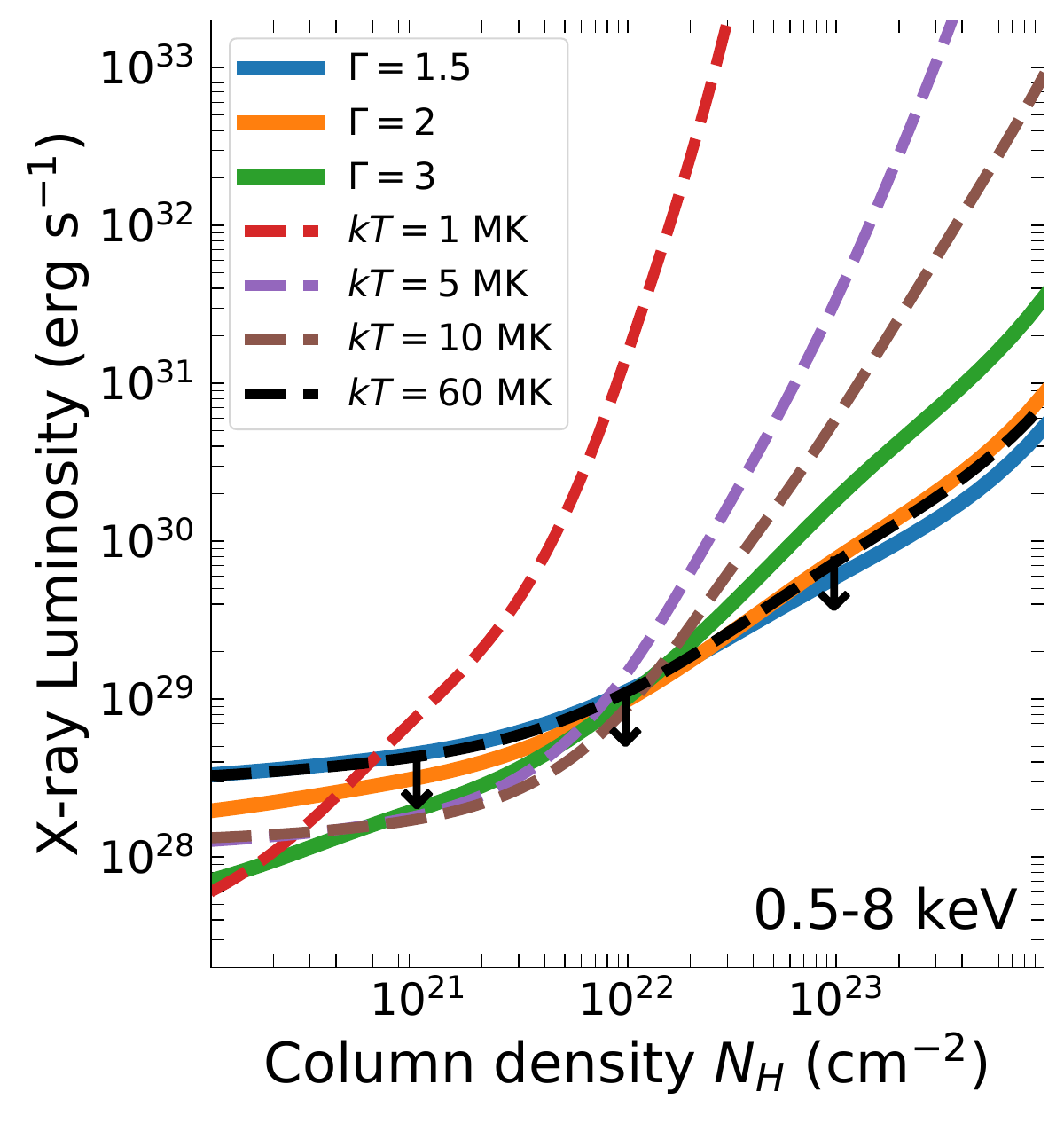}
      \caption{Upper limits on the unabsorbed X-ray luminosity ($0.5$\,$-$\,$8$ keV) of Betelgeuse and its companion as a function of the Hydrogen column density $N_H$ for a variety of model parameters. Solid lines refer to an absorbed powerlaw with photon index $\Gamma$\,$=$\,$\{1.5,2,3\}$, and dashed lines refer to a thermal APEC model with temperature $kT$\,$=$\,$\{1,5,10,60\}$ MK. The downward arrows are added to a single line to clarify it is an upper bound for that assumed model. All other lines are also upper bounds for the specific model, but arrows are not similarly shown to avoid crowding.
              }
    \label{fig:xraylimits}
\end{figure}

In order to determine the energy conversion factor (ECF) between count rate and unabsorbed flux, we next ran the \texttt{srcflux} command to generate the combined spectral files through the use of the \texttt{specextract} and \texttt{combine\_spectra} commands. Using the resulting auxiliary response file (ARF) and response matrix file (RMF), we ran the \texttt{modelflux} command over a grid of models to determine the unabsorbed flux ECF for each model. We consider two spectral models: \textit{i}) an absorbed powerlaw model ($\texttt{tbabs*pow}$) with photon index $\Gamma$ varied between $1$\,$-$\,$3.5$ and \textit{ii}) an absorbed thermal Astrophysical Plasma Emission Code \citep[APEC,][]{SmithAPEC2001} model ($\texttt{tbabs*apec}$) with temperature $kT$ between $0.01$\,$-$\,$5$ keV ($\sim$\,$0.1$\,$-$\,$60$ MK; \citealt[e.g.,][]{Getman2005COUP0,Preibisch2005,Preibisch2005Review}). We applied the abundance table from \citet{Wilms2000}, the photoelectric absorption cross-sections presented by \citet{Verner1996}, and adopted the default abundance for the APEC model given the near-solar abundance of Betelgeuse \citep{Ramirez2000}. For both models we consider a range of hydrogen column density $N_H$ between $10^{20-24}$ cm$^{-2}$ (but see \ref{sec:nHassumption} for further discussion). For each set of model parameters we computed the unabsorbed ECF between the $0.1$\,$-$\,$10$ keV count rate to the standard $0.5$\,$-$\,$8$ keV energy range (see Table \ref{tab: xrayresults}). All fluxes and luminosities in this manuscript are reported in the $0.5$\,$-$\,$8$ keV energy range. The results are shown in Figure \ref{fig:xraylimits}.

\begin{table*}[ht]
\centering
\caption{The merged $3\sigma$ upper limits on an X-ray source at the location of Betelgeuse based on our new \textit{Chandra} observations. The count rate is reported in the ``wide''  $0.1$\,$-$\,$10$ keV energy range (see Figure \ref{fig:fc}). We present the unabsorbed flux and unabsorbed luminosity limits (both in $0.5$\,$-$\,$8$ keV) for our fiducial APEC and powerlaw (PL) models, and our three estimates of hydrogen column density $N_H$, adopting a temperature of $kT$\,$=$\,$10$ MK (0.86 keV), and photon index of $\Gamma$\,$=$\,$2$. 
The list of all observations is shown in Table \ref{tab: observationsXray}.}

\label{tab: xrayresults}
\begin{tabular}{ccccccc}
\hline\hline
 \textbf{Instrument} &   \textbf{Exposure} &  \textbf{Count Rate}  & \textbf{$N_H$} &  \textbf{Model} & \textbf{Flux} & \textbf{Luminosity} \\
  &  \textbf{(ks)} & \textbf{(cts s$^{-1}$)} & \textbf{cm$^{-2}$} & & \textbf{(erg cm$^{-2}$ s$^{-1}$)} & \textbf{(erg  s$^{-1}$)}  \\
  & & \textbf{($0.1$\,$-$\,$10$ keV)} & & \textbf{($0.5$\,$-$\,$8$ keV)} & \textbf{($0.5$\,$-$\,$8$ keV)} \\
\hline
\hline
\textit{Chandra}/HRC-I & 41.85  & $<3.7\times10^{-4}$ & $6.2\times10^{21}$ & APEC & $<1.3\times10^{-14}$ & $<4.4\times10^{28}$  \\ 
... & ... & ... & ... & PL & $<2.0\times10^{-14}$ & $<6.9\times10^{28}$  \\ 
... & ...  & ... & $5.7\times10^{22}$ & APEC & $<5.3\times10^{-13}$ & $<1.8\times10^{30}$  \\ 
... & ... & ... & ... & PL & $<1.3\times10^{-13}$ & $<4.5\times10^{29}$  \\ 
... & ...  & ... & $3\times10^{23}$ & APEC & $<1.7\times10^{-11}$ & $<5.6\times10^{31}$  \\ 
... & ... & ... & ... & PL & $<5.8\times10^{-13}$ & $<2.0\times10^{30}$  \\ 
\hline \hline

\end{tabular}
\end{table*}

\section{Results and Discussion}\label{sec:results}

\subsection{X-ray Absorption from the Surrounding Environment}
\label{sec:nHassumption}

Due to its evolutionary stage as a red supergiant, Betelgeuse is continuously undergoing mass loss, releasing dust and gas in the form of stellar wind. Soft X-ray photons suffer from photoelectric absorption by these intervening particles, including gas, molecules, and grains \citep[see, e.g.,][]{Wilms2000}. The total amount of X-ray absorption is generally measured in terms of the amount of neutral hydrogen along a line-of-sight, which is usually measured in terms of a column density $N_H$. While hydrogen is not the main cause of X-ray absorption, which is usually dominated by more abundant heavier elements with more available line transitions, it is a general measure of the total amount of material between the source and the observer \citep[for a detailed discussion see, e.g.,][]{Wilms2000}. The relative abundance of elements along the line-of-sight is a key input to determine the total amount of X-ray absorption. As Betelgeuse displays a near solar abundance \citep{Ramirez2000}, we assume the same elemental abundance for the absorbing particles. 

The main source of uncertainty in constraining the intrinsic X-ray luminosity (as opposed to the absorbed luminosity) of either Betelgeuse or its companion lies in the total value of neutral hydrogen $N_H$ that is impacting our measurements. This includes the sum of multiple components, but we focus on the contribution of the Galactic interstellar medium (ISM) and the dense wind of Betelgeuse. Due to its very nearby distance (168 pc; \citealt{Joyce2020}), the interstellar medium is an extremely subdominant component. In fact, the total hydrogen column density inferred for this line-of-sight, for the entire Milky Way, is $4\times10^{21}$ cm$^{-2}$ \citep{Willingale2013}, and the value out to the distance of Betelgeuse will be significantly less. In Figure \ref{fig:xraylimits}, we computed a range of limits to the intrinsic X-ray luminosity over a wide range of $N_H$ values between $10^{20}$ cm$^{-2}$ and $10^{24}$ cm$^{-2}$. Now, we will more precisely estimate the $N_H$ density. 

First we examine approximations of the $N_H$ density from the literature. In a previous analysis on \textit{Chandra} data of Betelgeuse, \citet{Posson-Brown-2006} adopted an estimate of $6.2\times10^{21}$ cm$^{-2}$ for the hydrogen column density of the circumstellar shell surrounding Betelgeuse \citep{Hagen1978}. However, they note that much of the shell is likely to be composed of dust and grains, as opposed to entirely neutral atoms, which would lead to a smaller value of $N_H$ actually absorbing X-ray photons. 

Recently, \citet{Dent2024} detected Rydberg transition emission lines from Betelgeuse for the first time. A neutral hydrogen gas density above the photosphere as a function of radius from the star can be inferred from these lines. By integrating the radial number density profile (cm$^{-3}$) in Figure 3 of \citet{Dent2024} along our line of sight for the companion at quadrature (using an Abel transform to account for the geometry of the system and our line of sight to the companion, not Betelgeuse), from a radius of $\sim2.5$R$_{*}$ \citep{Goldberg2024} where R$_{*}$ = 764R$_{\odot}$, we estimate a $N_H$ column density of $\sim$3$\times10^{23}$ cm$^{-2}$.

Next, we compute our own estimate of the hydrogen column density due to the wind of Betelgeuse at the position of the companion star. As an approximation we assume that Betelgeuse's wind is constant and has an isotropic density distribution of $\rho(r)$\,$=$\,$\dot{M}_\textrm{w}/(4\pi v_\textrm{w} r^2)$ where $\dot{M}_\textrm{w}$ is the wind's mass loss, $v_\textrm{w}$ is the wind's velocity, and $r$ is the radius from Betelgeuse. We integrate this density profile from the predicted position of the companion, a distance $R$\,$=$\,$1850 R_\odot$ from Betelgeuse \citep[corresponding to $\approx$ 2.5 Betelgeuse radii;][]{Goldberg2024}, out to infinity along the line-of-sight at quadrature (again with an Abel transform, which analytically resolves to a factor of $\frac{\pi}{2}$). By adopting a hydrogen mass fraction of $X$\,$=$\,$0.7$ for the wind \citep[e.g.,][]{AndersGrevesse1989}, we convert the total particle column density to a hydrogen column density $N_H$. Adopting values of $\dot{M}_\textrm{w}$\,$\approx$\,$2\times10^{-6} M_\odot$ yr$^{-1}$ and $v_\textrm{w}$\,$\approx$\,$9$ km s$^{-1}$ \citep[e.g.,][]{Harper2001,Mauron2011,LeBertre2012,Dolan.M.2016.BetelEvo}, we find $N_H$\,$\approx$\,$5.7\times10^{22}$ cm$^{-2}$ for the likely hydrogen column density between us and the companion due to the wind of Betelgeuse, with a strong caveat that the 1/r$^{2}$ approximation assumes a uniform density of wind mass loss, while in reality the winds of RSGs (and Betelgeuse specifically) are known to be clumpy \citep{Mauron1997,Smith2009,Decin2012,Humphreys2022}. We do note that the likelihood of an \emph{extreme} wind clump along the line-of-sight during the time of our observation would be unlikely, given the lack of any significant optical dimming timed with our observation \citep{BetelHST2025}, whereas this was likely the case \citep{Montarges2021,Dupree2022} for previous \textit{Chandra} observations (where the companion was also eclipsed) during the Great Dimming period \citep{Kashyap-2020}.

As a final check, we consider the standard relation between optical extinction and hydrogen column density in our Galaxy \citep{Predehl1995,Guver2009}, applying most generally to the interstellar medium. Using the visual extinction ($A_V$\,$=$\,$0.65$ mag; \citealt{Montarges2021}) measured for Betelgeuse (containing contributions from the interstellar medium out to 168 pc and the contribution from its surrounding environment and wind), we find an estimate of $N_H$\,$\approx$\,$1.4\times10^{21}$ cm$^{-2}$. It is not a given that the surrounding environment and wind of Betelgeuse lie on this relation as the total Hydrogen column in the star's circumstellar environment likely contributes to X-ray absorption in addition to the dust that contributes to the measured visual extinction. However, if the standard $A_V$\,$-$\,$N_H$ relation holds, it would show that most of our estimates, as expected, are overconservative and that our conclusions should become more robust. Based on Figure \ref{fig:xraylimits}, for the majority of models this smaller $N_\textrm{H}$ value would decrease the upper limits be a factor of a few.

\subsection{Unabsorbed X-ray flux and luminosity limits}

The intrinsic constraint on the X-ray flux or luminosity depends strongly on the assumed hydrogen column density, which modifies the conversion between counts and flux (the ECF) for a given spectral model (see \S \ref{sec: chandraanalysis}). In Table \ref{tab: xrayresults}, we report the upper limits to the unabsorbed flux and luminosity in $0.5$\,$-$\,$8$ keV for the three estimates of $N_H$ outlined in \S \ref{sec:nHassumption} assuming two fiducial models for the X-ray spectrum of $\alpha$ Ori B. The full impact of the possible range of $N_H$ values is shown in Figure \ref{fig:xraylimits}. We have assumed a distance of 168 pc \citep{Joyce2020}. The uncertainty on the distance of Betelgeuse is minimal ($168^{+27}_{-15}$ pc; see Figure 10 of \citealt{Joyce2020}) and has a significantly smaller impact than the uncertainty in $N_H$.

We adopt two fiducial spectral models: a thermal plasma model and an absorbed powerlaw model (see \S \ref{sec: chandraanalysis} for further discussion). 
For the thermal plasma  (APEC) model with $kT$\,$=$\,$10$ MK (0.86 keV), the limit is $<$\,$1.7\times10^{-11}$ erg cm$^{-2}$ s$^{-1}$ and $L_X$\,$\lesssim$\,$5.6\times10^{31}$ erg s$^{-1}$ for the more conservative $N_H =3\times10^{23}$ cm$^{-2}$ column density, and is $<$\,$5.3\times10^{-13}$ erg cm$^{-2}$ s$^{-1}$ and $L_X$\,$\lesssim$\,$1.8\times10^{30}$ erg s$^{-1}$ for the $N_H = 5.7\times10^{22}$ cm$^{-2}$ estimate. For the powerlaw model with $\Gamma$\,$=$\,$2$, the limit is $<$\,$5.8\times10^{-13}$ erg cm$^{-2}$ s$^{-1}$ and $L_X$\,$\lesssim$\,$2.0\times10^{30}$ erg s$^{-1}$ for the $N_H =3\times10^{23}$ cm$^{-2}$ estimate, and is $<$\,$1.3\times10^{-13}$ erg cm$^{-2}$ s$^{-1}$ and $L_X$\,$\lesssim$\,$4.5\times10^{29}$ erg s$^{-1}$ for the $N_H = 5.7\times10^{22}$ cm$^{-2}$ estimate. These limits conservatively constrain the allowed range of intrinsic luminosity for $\alpha$ Ori B. In what follows we discuss the implications of these limits and their ability to constrain the nature of $\alpha$ Ori B.

\subsection{A White Dwarf Companion}

A white dwarf (WD) falls within the mass range allowed for the proposed companion \citep{Goldberg2024,MacLeod2024}. While this is difficult to produce through standard evolutionary channels, we consider the possibility that Betelgeuse is a symbiotic star \citep[e.g.,][]{Kenyon1990,Munari2019}, where the WD accretes from Betelgeuse's wind \citep[e.g., Bondi-Hoyle accretion;][]{BondiHoyle1944}. In general, symbiotic stars are defined as a WD accreting from a late-type giant star. As Betelgeuse is a supergiant, its mass loss is significantly higher than expected from a typical lower mass giant. In fact, there are no observed supergiant symbiotic stars in the Milky Way. 

The lack of observed analogous systems may be due to the difficulty in producing them through standard binary evolution. To investigate this possibility, we ran a series of simulations with COSMIC, a rapid binary population synthesis code \citep{Breivik2020}, with initial primary masses varying between $8-15\,\rm{M_{\odot}}$, mass ratios between $0.5-1.0$, and orbital periods between $10-1,000$ days. We also investigated different assumptions for accretion efficiency during stable mass transfer since a mass ratio reversal would be required to form a WD companion to a Betelgeuse-mass star. Out of $10,000$ initial binaries, we were unable to produce a Betelgeuse-like binary with a WD companion with the parameters inferred by \cite{Goldberg2024, MacLeod2024}. The primary outcomes were either the onset of a common envelope evolution which shrinks the binary or the production of a NS companion (see Section~\ref{sec:neutronstar}). We note, though, that COSMIC does not account for effects on stellar evolution during mass loss or mass gain, so future studies could investigate this possibility with a detailed stellar evolution study.

In general, Betelgeuse's current mass suggests that in order to have a WD companion at present, there must have been a significant donation of mass from the WD progenitor. This would require that the binary begin at a much shorter period than at present and in the absence of WD natal kicks, which should be limited to a few $\rm{km/s}$ at best \cite[e.g.][]{Fellhauer2003, Heyl2007}, would not allow for the presently observed binary properties. If the mass exchange was instead initiated in a wider orbit, it is extremely unlikely that the core mass of the WD progenitor would be low enough to not produce a NS while still being able to donate enough envelope mass to Betelgeuse to lead to its presently measured mass. If no mass exchange occurs at all, the lifetime of \emph{any} WD progenitor ($>$50 Myr) far exceeds the age of Betelgeuse ($\sim$10 Myr) and is thus ruled out. However, below we still comment on the X-ray luminosity of such systems in the context of our observational limits. We note that such a system could potentially be produced through alternative formation mechanism, such as a hierarchical triple system with a merger leading to a binary.

Symbiotic stars are strong X-ray emitters \citep[e.g.,][]{Eze2011,Luna2013} with typical X-ray luminosities of $10^{30-34}$ erg s$^{-1}$ \citep[e.g.,][]{Luna2013,Mukai2017,Lima2024} with plasma temperatures in the range of a few keV. Based on their observed X-ray spectral properties they have been classified into multiple categories \citep{Muerset1997,Luna2013}. The X-rays are generally thought to be produced at the wind collision boundary between the WD and and the donor star ($\beta$-type) or by the boundary layer of the accretion disk ($\delta$-type; e.g., T CrB). 
WD symbiotic binaries typically display either hard, single component spectra (usually described by an APEC model), and, in some cases, also the addition of a soft component at $<$\,$2$ keV \citep[$\alpha$-type;][]{Luna2013}. 

The significant mass loss of Betelgeuse, around $10^{-6}M_\odot$ yr$^{-1}$, would likely form an accretion disk around the WD \citep[e.g.,][]{LivioWarner1984} leading to a supergiant analog of a $\delta$-type symbiotic star. 
If the innermost boundary layer of the accretion disk is optically thin \citep[see][]{NarayanPopham1993} the source will be luminous in X-rays. 
The accretion rates of WD symbiotic systems are generally within the range of $10^{-9}$ to $10^{-8}M_\odot\,\textrm{yr}^{-1}$, depending on the WD mass, and can be inferred under the assumption that $\dot{M}_\textrm{acc}$\,$=$\,$2L_X R_\textrm{WD}/(G M_\textrm{WD})$. 
The expectation for Bondi-Hoyle accretion \citep{BondiHoyle1944} is consistent with these inferred values, yielding $\dot{M}_\textrm{acc}$\,$\approx$\,$10^{-8}M_\odot\,\textrm{yr}^{-1}\,(M_\textrm{WD}/0.6M_\odot)^2\,(7\,\textrm{km}\,\textrm{s}^{-1}/v_\infty)^3$ where $v_\infty$ is the relative velocity of the wind \citep[see][]{Luna2013}. We note that this is significantly smaller than the mass loss rate of Betelgeuse which would require $\dot{M}_\textrm{acc}/\dot{M}_\textrm{w}$\,$\lesssim$\,$10^{-2}$ to reach similar values and thus similar expected luminosities. It is also worth noting that higher accretion rates may lead to an optically thick inner boundary layer that may suppress X-ray emission \cite[see, e.g.,][]{NarayanPopham1993}. 

The observed luminosities of known symbiotic stars \citep[$10^{30-34}$ erg s$^{-1}$; see, e.g.,][]{Luna2013} exceed our upper limit on the underlying X-ray luminosity ($\lesssim$\,$10^{30}$ erg s$^{-1}$), by multiple orders of magnitude on average. Therefore, we can strongly disfavor a WD as the companion to Betelgeuse for a wide range of assumptions for the absorbing hydrogen column density (Figure \ref{fig:xraylimits}). This conclusion is supported by stellar population synthesis, which would require an extremely peculiar formation channel to create a WD - supergiant system.

\subsection{A Neutron Star Companion}\label{sec:neutronstar}

Similarly, a neutron star (NS) falls within the range of allowed masses for Betelgeuse's companion \citep{Goldberg2024,MacLeod2024}. A NS companion to Betelgeuse would also be supported by common evolutionary channels. 
As most high-mass stars are born in binary systems \citep[e.g.,][]{Sana2012, deMink2013, Toonen2018, Toonen2020, Toonen2022, Offner2023}, and the end stage of their lives result in a compact object (either a NS or a black hole), it would make sense to have a situation where a NS is in a binary with an evolved high-mass star. In this case, Betelgeuse would be an evolved stage of a high-mass X-ray binary \citep[HMXB; see, e.g.,][]{Tauris2017}. In that scenario, the progenitor to a NS born at the same time as Betelgeuse was most likely a more massive star than Betelgeuse in order to die earlier (within $\sim$10 Myr) in a supernova explosion.

The supernova explosion of this previously-high-mass companion (resulting in its transformation to a NS) would also likely produce observable changes in the system's velocity due to the natal kick at the NS's birth \citep[e.g.][]{Eldridge2011, Zapartas2021}, which may widen the orbit without breaking it. This would especially be the case in the event that the progenitor to the NS was a highly stripped star (due to prior interaction of the binary system) where the supernova ejecta mass is comparable to the NS mass. If $\alpha$Ori~B were the compact remnant of this past explosion, it could confirm a supernova runaway as the explanation for the high $\sim$20 km s$^{-1}$ proper motion of Betelgeuse \citep[e.g.][]{vanLeeuwen2007,Harper2008, Harper2017}. Cluster ejection via dynamical capture of a stellar-mass companion is a possible alternative explanation; however Betelgeuse's proper motion has not yet been traced back to a specific cluster. Additionally, almost all ejections happen within 1 Myr of the cluster formation \citep{Oh2016}, but the age of Betelgeuse and all nearby clusters and OB associations is much older than 1 Myr.

As Betelgeuse's radius ($\sim$764R$_\odot$) is less than it's Roche radius ($R_\textrm{Roche}$\,$\approx$\,$1100R_\odot$), the NS companion would be accreting from the wind of Betelgeuse. Analogously to symbiotic stars, a NS accreting from the wind of a late-type giant star is referred to as a symbiotic X-ray binary (SyXRB).
They have typical X-ray luminosities in the range of $10^{32-36}$ erg s$^{-1}$ \citep[e.g.,][]{Yungelson2019,De2022a,De2022b}. Their X-ray spectra are quite hard with typical X-ray photon indices in the range of $\Gamma$\,$\approx$\,$1$ to $2$. These are rare systems with only around a dozen known in our Galaxy \citep{Yungelson2019}, which is likely due to their relatively uncommon formation channels. In the past few years, two red supergiant SyXRBs were discovered \citep[4U 1954+31 and SWIFT J0850.8-4219;][]{Hinkle2020,De2024}, with a third candidate requiring spectroscopic confirmation \citep[XID 6592;][]{Gottlieb2020}. These sources are essentially evolved HMXBs (i.e., supergiant+NS in a binary). If the inferred companion to Betelgeuse is a NS, then these systems are direct analogs. The red supergiant SyXRBs display X-ray luminosities of $\sim$\,$10^{35}$ erg s$^{-1}$ \citep[][]{Hinkle2020,De2024}
, significantly more luminous than our X-ray constraints. In addition, they are hard X-ray emitters (photon index $\Gamma$\,$<$\,$1$; \citealt{De2024}) and therefore less impacted by absorption due to the surrounding environment.

We can estimate the theoretical X-ray luminosity from the Betelgeuse system under the standard assumption that the NS with mass $M_\textrm{NS}$ accretes from Betelgeuse's wind with mass loss rate $\dot{M}_\textrm{w}$ and velocity $v_\textrm{w}$ \citep{BondiHoyle1944}:
\begin{align}
    \dot{M}_\textrm{acc} = 
    \frac{5 G^2 M_\textrm{NS}^2\dot{M}_\textrm{w}}{8 v_\textrm{rel}^3v_\textrm{w} a^2},
\end{align} 
which leads to an X-ray luminosity of 
\begin{equation}
    L_X = \eta \dot{M}_\textrm{acc} c^2,
\end{equation} 
where the radiative efficiency $\eta$ is assumed to be 10\%. The orbital velocity of the NS companion is taken to be $v_\textrm{orb}$\,$=$\,$2\pi a/T_\textrm{orb}$\,$\approx$\,$43$ km s$^{-1}$ where $T_\textrm{orb}$\,$=$\,$2170$ d and $a$\,$=$\,$(1850 \pm 70)R_\odot$ \citep{Goldberg2024}. A similar value is obtained from $v_\textrm{orb}$\,$=$\,$\sqrt{G(M_\textrm{NS}+M_\textrm{Betel})/a}$\,$\approx$\,$45$ km s$^{-1}$ where we assume a neutron star mass of $M_\textrm{NS}$\,$=$\,$1.4 M_\odot$ and adopt the mass of Betelgeuse $M_\textrm{Betel}$\,$=$\,$(18\pm 1) M_\odot$ \citep{Joyce2020}. 
As the velocity of Betelgeuse's wind is only $\sim$\,$15$ km s$^{-1}$, the relative velocity $v_\textrm{rel}^2$\,$=$\,$v_\textrm{w}^2+v_\textrm{orb}^2$ is dominated by the orbital velocity such that $v_\textrm{rel}$\,$\approx$\,$v_\textrm{orb}$. Inserting these numbers, we find: 
\begin{align}
   \frac{\dot{M}_\textrm{acc}}{\dot{M}_\textrm{w}} 
    = 0.01 
    \left(\frac{M_\textrm{NS,*}}{a_\textrm{sep}}\right)^2
    \left(\frac{v_\textrm{w}}{15\,\textrm{km\,s}^{-1}}\right)^{-1} \left(\frac{v_\textrm{orb}}{45\,\textrm{km\,s}^{-1}}\right)^{-3},
\end{align} 
where we have applied $a_\textrm{sep}$\,$=$\,$a/(1850 R_\odot)$ and $M_\textrm{NS,*}$\,$=$ $M_\textrm{NS}/(1.4M_\odot)$. For an assumed mass loss rate of $\dot{M}_\textrm{w}$\,$\approx$\,$10^{-6}M_\odot\,\textrm{yr}^{-1}$ this yields an X-ray luminosity $L_X$\,$\approx$\,$6\times10^{37}$ erg s$^{-1}$. Even if $\dot{M}_\mathrm{w}$ or efficiency $\eta$ were vastly lower (e.g., $10^{5-6}$ times), we would still expect X-ray luminosities well in excess of $L_X>10^{32}$ erg s$^{-1}$, which is over two orders of magnitude greater than our X-ray limits in Table \ref{tab: xrayresults}. 

Theoretical expectations for supergiant+NS systems assuming Bondi-Hoyle accretion \citep{BondiHoyle1944} lead to predicted X-ray luminosities as high as $\sim$\,$10^{37}$ erg s$^{-1}$ \citep[see][for a larger discussion]{De2024}. Interestingly, these values are nearly a factor of 100 higher than the luminosities observed from confirmed supergiant+NS systems ($\sim$\,$10^{35}$ erg s$^{-1}$). There have been two main explanations proposed for the lower observed X-ray luminosities of the two known supergiant -  neutron star systems \citep{Hinkle2020,De2024}: \textit{i}) that the NS is at an extremely large separation from the supergiant  where some of the assumptions of spherical Bondi-Hoyle accretion start to break down \citep{Hinkle2020,De2024} and \textit{ii}) that the NS is in what is known as a `propeller state' that decreases the accretion rate by stopping the wind at the magnetospheric radius \citep{De2024}. 
\citet{De2024} have argued that large separations do not provide the most natural explanation. Regardless, in our case, the inferred separation of Betelgeuse and the companion is too close, $a$\,$=$\,$(1850 \pm 70)R_\odot$ \citep{Goldberg2024}, to apply a large separation to lead to a lower accretion rate and, therefore, a lower luminosity. The added difficulty in hiding such a supergiant - NS system in our case is that we would have to drop the X-ray luminosity by significantly more (nearly by a factor of 10 million, rather than a factor of 100) to be consistent with our observed X-ray limits from \textit{Chandra} (Figure \ref{fig:xraylimits} and Table \ref{tab: xrayresults}). As such, a neutron star companion is ruled out within our present theoretical and observational understanding.

\subsection{A Young Stellar Object}

\begin{figure*}
   \centering
  \includegraphics[width=0.9\textwidth]{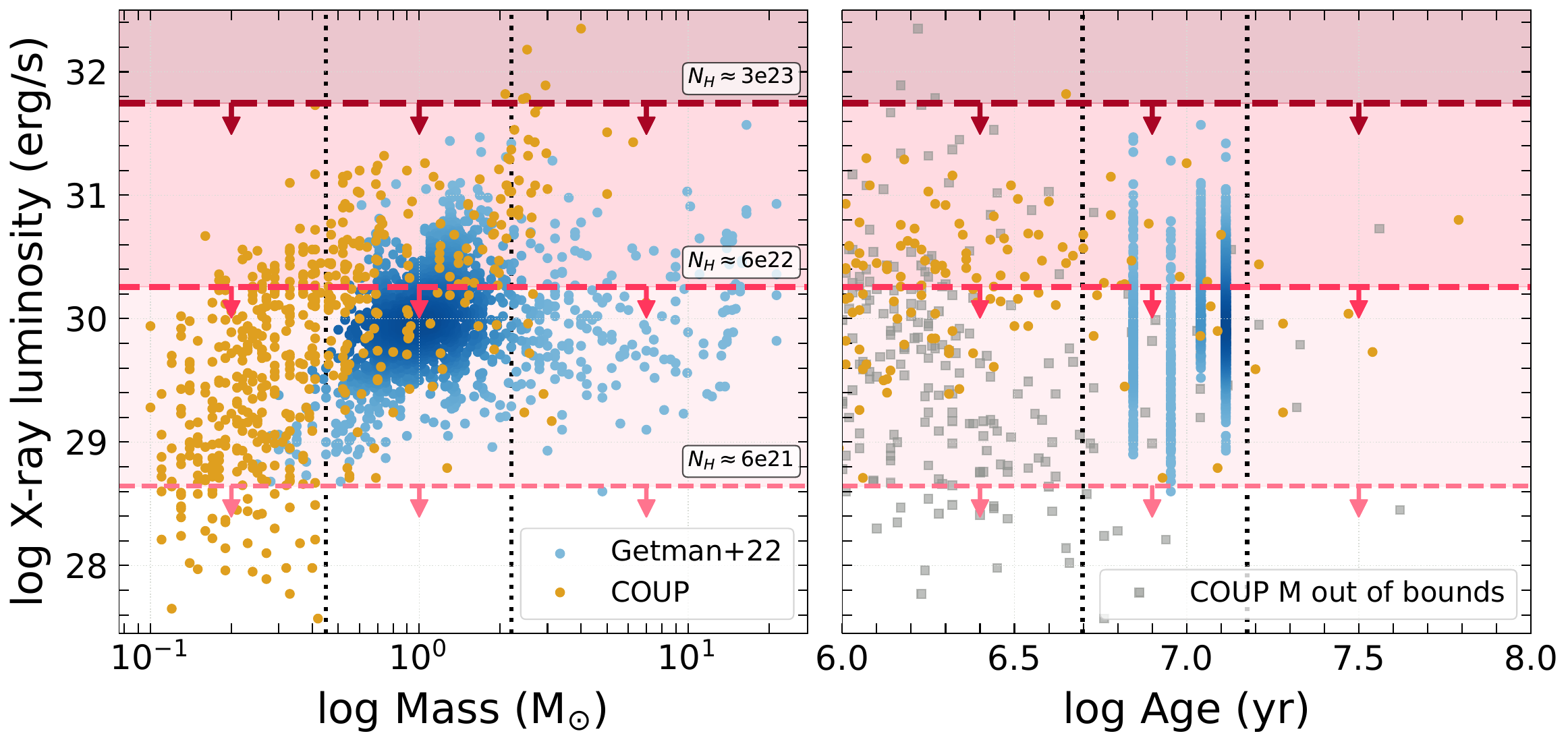}
      \caption{Unabsorbed X-ray luminosity ($0.5$\,$-$\,$8$ keV) versus stellar mass (left panel) and age (right panel). YSOs from \citet{Getman2005COUP0} and \citet[][only clusters with ages between $5$\,$-$\,$15$ Myr are considered]{Getman2022} are plotted as gold and blue circles, respectively. In both panels, darker shading indicates a higher density of sources for the \citet{Getman2022} YSOs, and in the right panel \citet{Getman2005COUP0} YSOs with masses outside the estimated range of the companion are plotted as grey squares to aid in viewing. Our lower limits on the unabsorbed X-ray luminosity of the system, assuming an APEC thermal plasma model of $kT$\,$=$\,$10$ MK are indicated with dashed red lines, of increasing thickness and darkness corresponding to our estimate of the neutral Hydrogen column density of increasing denseness. Dotted black lines indicate the mass estimate for the companion ($0.45\leq\frac{\mathrm{M}}{\mathrm{M}_\odot}\leq2.2$) and the conservative age estimate for Betelgeuse ($5$\,$-$\,$15$ Myr).
               }
    \label{fig:lim-YSO-result}
\end{figure*}

For the case of a wide natal binary with a large mass ratio, given the age of Betelgeuse ($\sim$10 Myr \citealt{Joyce2020}), the pre-MS lifetime of low-mass stellar objects means that the companion would likely be a Young Stellar Object (YSO) \citep{Tognelli2011yso,Baraffe2015yso,Haemmerle2019yso,Amard2019yso}, though a very young MS star is possible in the case of a 2M$_\odot$ companion. We note that this large mass ratio is not impossible at this separation, and studies \citep{Rizzuto2013, Moe2017, Offner2023} have found that binaries with wider separations have mass ratios skewed towards smaller companions, favoring $q<0.2$ at $\approx$10AU.

YSOs including those with similar ages to Betelgeuse and within the mass range allowed for the companion \citep{Goldberg2024,MacLeod2024}, have had their X-ray emissions extensively studied for decades (see \citealt{Feigelson1999Review,Preibisch2005Review} for reviews). YSOs demonstrate strong X-ray activity throughout their evolution, remaining luminous while young ($\lessapprox$\,$10$ Myr) and then declining until reaching the zero-age main-sequence. They are X-ray bright due to strong magnetic dynamo processes leading to coronal ejections and flares \citep{Feigelson1999Review,Preibisch2005,Preibisch2005Review,Getman2022}. This X-ray activity is ubiquitous across mass; \citet{Preibisch2005} found no evidence for a population of low-mass YSOs with low or no X-ray activity. 

Further, the Chandra Orion Ultradeep Project \citep[COUP;][]{Flaccomio2003COUP1,Flaccomio2003COUP2,Getman2005COUP0}, took extremely deep observations of star forming regions in Orion using \textit{Chandra}, surveyed over 1600 YSOs and characterized their X-ray emission. The targets surveyed ranged in ages from $\sim$\,$0.1$\,$-$\,$10$ Myr. In an analysis of $\sim$\,$600$ lightly absorbed and well-characterized YSOs, \citet{Preibisch2005} found a positive trend in X-ray luminosity with mass, but a \emph{large} scatter (see Figure 3 of \citealt{Preibisch2005}). This scatter is attributed to the influence of accretion, as accreting YSOs do not follow a luminosity-mass relationship as strongly as non-accreting YSOs. Accreting YSOs also have, in general, slightly lower X-ray luminosities than non-accreting YSOs. We note that $\bbud$ is likely \emph{not} accreting because it likely no longer has a disk. Circumstellar disks can be photoevaporated by UV radiation from nearby OB stars \citep{Johnstone1998,Adams2004,Andrews2020}, and Betelgeuse was a hot O-star for it's MS lifetime. Also, \citet{Kraus2012} finds that for YSO binary systems with close separations (r$\lessapprox$40\,AU -- the separation of $\bbud$ is only $\sim$9AU), 2/3rds of these systems disperse their disks within 1\,Myr of formation.

X-ray luminosities for YSOs with masses in the range of values predicted for the companion (adopting the broadest range of values from \citealt{MacLeod2024,Goldberg2024}),  M\,$=$\,$0.45$\,$-$\,$2.2\mathrm{M}\odot$, are $\sim$\,$1\times10^{28}$ to $7\times10^{32}$ erg s$^{-1}$. \citet{Getman2005COUP0} observed 10 YSO clusters across a range of ages, and (for clusters with ages similar to that of Betelgeuse of $\sim$\,$10$ Myr; \citealt{Joyce2020}) found a similar range of X-ray luminosities. 

Therefore, we expect a YSO companion to Betelgeuse to similarly display bright X-ray emission, regardless of its exact mass. We have computed the limits on unabsorbed X-ray luminosity over a wide range of hot plasma temperatures, covering the range found in past \textit{Chandra} observations of YSOs \citep[e.g.,][]{Getman2005COUP0,Preibisch2005}. These limits are shown in Figure  \ref{fig:xraylimits} as dashed lines.

In Figure \ref{fig:lim-YSO-result}, we show the X-ray luminosities of of YSOs from \citet{Getman2005COUP0} (gold) and \citet{Getman2022} (blue circles) as a function of mass (left panel, $0.45$\,$-$\,$2.2\mathrm{M}\odot$ enclosed by black dotted lines) and age (right panel, $5$\,$-$\,$15$ Myr enclosed by black dotted lines). We show the upper limits on the X-ray luminosity of the companion from our \emph{Chandra} observations as three dashed red lines of increasing thickness and darkness, corresponding to the three estimate of $N_H$ column density of increasing denseness (\S\ref{sec:nHassumption}). X-ray luminosities above the highest line ($N_H = 3\times10^{23}$ cm$^{-2}$) are strongly excluded by our observations, though few examples of YSOs have X-ray luminosities this high. Taking the middle estimate from the wind mass loss of Betelgeuse instead ($N_H = 6\times10^{22}$ cm$^{-2}$), of the 176 \citet{Getman2005COUP0} and 1612 \citet{Getman2022} YSOs with M\,$=$\,$0.45$\,$-$\,$2.2\mathrm{M}\odot$, 59\% of the \citet{Getman2005COUP0} sources and 25\% of the \citet{Getman2022} sources are excluded. There is a slight trend towards higher masses for excluded YSOs -- the mean mass in M$_{\odot}$ of excluded YSOs is 0.76 \citep{Getman2005COUP0} and 1.0 \citep{Getman2022}, compared to 1.1 \citep{Getman2005COUP0} and 1.2 \citep{Getman2022} for allowed YSOs. Thus we find that there is a range of parameter space consistent with the mass estimates of $\bbud$ which would fall below our detection threshold.

\subsection{Constraints on the Surface Flux of Betelgeuse}
\label{sec:betelsurfconstraints}

While our latest set of \textit{Chandra} observations are optimally timed (Figure \ref{fig:timing}) to constrain the companion of Betelgeuse, we can combine all available \textit{Chandra} data to set a deeper limit on the surface flux of Betelgeuse itself \citep[e.g.,][]{Posson-Brown-2006}. We followed the same steps outlined in \S \ref{sec: chandraanalysis} and analyzed 59.8 ks of HRC-I data by combining the latest set of data (Table \ref{tab: observationsXray}; PI: O'Grady) with past calibration observations reported by \citet{Posson-Brown-2006}, and the previous DDT observations presented by \citet{Kashyap-2020}, see \S \ref{sec: archival} and Table \ref{tab: archival}. The merged count rate limit from \texttt{srcflux} is $<2.1\times10^{-4}$ cts s$^{-1}$ in the $0.1$\,$-$\,$10$ keV band, nearly a factor of 2 more sensitive than the limits on the companion. A similar count rate limit is computed in the filtered $0.5$\,$-$\,$8$ keV range, where in the merged image (using \texttt{merge\_obs}) we identify 7 photons in our 1.5\arcsec\, circular source region with 6.7 expected background photons. We note that these limits are independent of proper motion corrections \citep[e.g.,][]{Harper2017} to the source aperture as a function of observing epoch between December 2001 and December 2024 (see Tables \ref{tab: observationsXray} and \ref{tab: archival}). 

Furthermore, the ECF values derived using all available 59.8 ks of HRC-I data differ by only 5\% compared to the new 41.85 ks of data. Assuming an APEC model with temperature 10 MK and $N_H$\,$=$\,$10^{22}$ cm$^{-2}$ yields a limit of $<2.1\times10^{-14}$ erg cm$^{-2}$ s$^{-1}$ and $<7.2\times10^{28}$ erg s$^{-1}$ in $0.5$\,$-$\,$8$ keV. For $N_H$\,$=$\,$10^{21}$ cm$^{-2}$ the limit is a factor of $10\times$ more sensitive. 

The corresponding limit to the ratio of X-ray luminosity to bolometric luminosity of Betegleuse is $L_X/L_\textrm{bol}$\,$\lesssim$\,$10^{-9}$ to $10^{-10}$, depending slightly on the variability of Betelgeuse. 
This confirms Betelgeuse as a dark supergiant \citep{Posson-Brown-2006}. As our limits are of similar depth to those presented in \citet{Posson-Brown-2006}, our constraints on the corona temperature of Betelgeuse are similar, and can therefore rule out coronal plasma with temperatures of $>$\,$1$ MK. Low temperature plasma, especially in the presence of a significant absorption column, cannot be excluded (see \citealt{Posson-Brown-2006} for details).

\section{Conclusions} 
\label{sec:conclusions}

We have presented our analysis of recent \textit{Chandra} observations of Betelgeuse in December 2024 during the period of maximum separation between Betelgeuse and its proposed companion. We do not detect an X-ray source at the proper motion corrected position of Betelgeuse and instead place $3\sigma$ upper limits on the underlying X-ray luminosity of roughly $L_X$\,$\lesssim$\,$10^{30-31}$ erg s$^{-1}$ ($\lesssim$\,$2.6\times10^{-3}L_\odot$ to $2.6\times10^{-4}L_\odot$) in $0.5$\,$-$\,$8$ keV for the expected hydrogen column densities $N_H$\,$=$\,$10^{22-23}$ cm$^{-2}$ from Betelgeuse's wind. This limit is robust to different model assumptions under this specific $N_H$ value. These X-ray limits exclude a compact object (e.g., white dwarf, neutron star) as the companion to Betelgeuse. While we therefore strongly favor a young stellar object as the companion, we cannot place strong constraints on its mass, though a sub-solar mass companion is slightly favored (see \citealt{BetelHST2025} for further discussion of complementary UV constraints from the \textit{Hubble Space Telescope}). Finally, we note that during the review process for this article, a possible direct-image detection of the companion was published \citep{Howell2025BB}. This work suggests that the companion is a stellar object with a mass of 1.4-2M$_\odot$, which is within the overall predicted mass range, and specifically consistent with our results that the companion cannot be a compact object.

\section*{Acknowledgements}
The authors thank Andrea Dupree, Saavik Ford, Yuri Levin, Morgan MacLeod, Barry McKernan, Koji Mukai, Adiv Paradise, Mathieu Renzo, and Leike van Son for useful discussions. A.O. also acknowledges the support of Jupiter O'Grady. The authors also thank the anonymous reviewer for a helpful and constructive referee report.

A.O. and B.O. gratefully acknowledge support from the McWilliams Postdoctoral Fellowship in the McWilliams Center for Cosmology and Astrophysics at Carnegie Mellon University. J.A.G. acknowledges a Flatiron Research Fellowship at the Flatiron Institute supported by the Simons Foundation. M.J. gratefully acknowledges funding of MATISSE: \textit{Measuring Ages Through Isochrones, Seismology, and Stellar Evolution}, awarded through the European Commission's Widening Fellowship. J.H. acknowledges support from NASA under award number 80GSFC21M0002. This project has received funding from the European Union's Horizon 2020 research and innovation programme. This research was supported by the `SeismoLab' KKP-137523 \'Elvonal grant and by the NKFIH excellence grant TKP2021-NKTA-64 of the Hungarian Research, Development and Innovation Office (NKFIH).  M.R.D. acknowledges support from the NSERC through grant RGPIN-2019-06186, the Ontario Ministry of Colleges and Universities through grant ER22-17-164, the Canada Research Chairs Program, and the Dunlap Institute at the University of Toronto.

This paper employs a list of Chandra datasets, obtained by the Chandra X-ray Observatory, contained in the Chandra Data Collection \dataset[doi:10.25574/cdc.424]{https://doi.org/10.25574/cdc.424}.
The scientific results reported in this article are based on observations made by the Chandra X-ray Observatory (CXO). This research has made use of data obtained from the Chandra Data Archive provided by the Chandra X-ray Center (CXC). This research has made use of software provided by the Chandra X-ray Center (CXC) in the application package \texttt{CIAO}. 
This research made use of NASA’s Astrophysics Data System Bibliographic Services, the \emph{pathfinder} framework \citep{Pathfinder} as well as of the SIMBAD database and the cross-match service operated at CDS, Strasbourg, France. 

\vspace{5mm}
\facilities{\textit{Chandra X-ray Observatory}}

\software{\texttt{CIAO} \citep{Ciao}}

\bibliographystyle{aasjournal}
\singlespace

\bibliography{RSG3D.bib, LSP_ADS.bib, BetelBuddy.bib, bib.bib, Joyce.bib}

\appendix
\restartappendixnumbering 
\section{Log of Observations}

Here we list all the \textit{Chandra} data of Betelgeuse analyzed in this work (see Tables \ref{tab: observationsXray} and \ref{tab: archival}). The data reported in Table \ref{tab: observationsXray} can be directly applied to constrain the companion properties as it is timed for the maximal separation (quadrature) from Betelgeuse \citep[see][]{Goldberg2024,MacLeod2024}, whereas the data in Table \ref{tab: archival} should only be used to constrain X-ray emission from Betelgeuse itself (see \S \ref{sec:betelsurfconstraints}). 

\begin{table*}[ht]
\centering
\caption{Log of the recent \textit{Chandra} X-ray observations (PI: A. O'Grady; Program: 25209014) of Betelgeuse used in this work. 
}
\label{tab: observationsXray}
\begin{tabular}{ccccc}
\hline\hline
\textbf{Start Time} & \textbf{Instrument} &  \textbf{ObsID} & \textbf{Exposure} \\
\textbf{(UT)} &  &    & \textbf{(ks)} \\
\hline
 2024-12-09 04:23:07 & \textit{Chandra}/HRC-I & 30590 & 4.91 \\ 
 2024-12-10 08:30:21	 & \textit{Chandra}/HRC-I & 30669 & 4.51 \\ 
 2024-12-11 09:05:51 & \textit{Chandra}/HRC-I & 30670 & 8.88 \\
 2024-12-11 23:57:12 & \textit{Chandra}/HRC-I & 30671 & 5.88 \\ 
 2024-12-12 23:56:37 & \textit{Chandra}/HRC-I & 30672  & 4.85  \\ 
 2024-12-13 18:57:59 & \textit{Chandra}/HRC-I & 30673 & 8.73  \\ 
 2024-12-15 13:47:42 & \textit{Chandra}/HRC-I & 30674 & 4.09   \\ 
  \hline
 \hline
    \multicolumn{4}{c}{\textbf{Merged Observation}}  \\
       \hline
... & \textit{Chandra}/HRC-I & ... & 41.85  \\ 
\hline\hline
\end{tabular}
\end{table*}

\begin{table*}[ht]
\centering
\caption{Log of the archival \textit{Chandra} X-ray observations  used in this work. References: [1] \citet{Posson-Brown-2006} [2] \citet{Kashyap-2020}
}
\label{tab: archival}
\begin{tabular}{cccccc}
\hline\hline
\textbf{Start Time} & \textbf{Instrument} &  \textbf{ObsID} & \textbf{Exposure} & \textbf{Reference} \\
\textbf{(UT)} &  &    & \textbf{(ks)} &  \\
\hline
 2001-12-07 08:53:15 & \textit{Chandra}/HRC-I & 2596 & 1.88 &  [1] \\ 
 2003-02-06 06:13:56	 & \textit{Chandra}/HRC-I & 3680 & 1.88  & [1]\\ 
 2004-02-02 09:52:24 & \textit{Chandra}/HRC-I & 5055  & 2.04 & [1]\\
 2005-02-02 18:33:00 & \textit{Chandra}/HRC-I & 5970 & 2.13 & [1]\\ 
 2020-02-17 19:16:29	 & \textit{Chandra}/HRC-I & 23147 & 5.09 & [2] \\ 
 2020-08-15 11:41:41 & \textit{Chandra}/HRC-I & 23148 & 4.98 & This work  \\ 
  \hline
 \hline
\end{tabular}
\end{table*}

\end{document}